\documentclass[useAMS,usenatbib]{mn2e}
\usepackage{graphicx,subfigure}
\usepackage{amsmath}
\usepackage{rotating}
\usepackage{color}
\usepackage{times}
\usepackage{mathtools}
\begin{document}

\def \apj {ApJ}
\def \apss {Ap{\&}SS}
\def \mnras {MNRAS}
\def \aj {AJ}
\def \araa {ARA\&A}
\def \aapr {A\&ARv}
\def \pasa {PASA}
\def \aap {AAP}
\def \aaps {AAPS}
\def \apjl {ApJL}
\def \apjs {ApJS}
\def \pasj {PASJ}
\def \nat {Nature}

%
%
%
%


\newcommand{\kpc}{{\rm\,kpc}}
\newcommand{\kms}{{\rm\,km\,s^{-1} }}
\newcommand{\msun}{{\rm\,M_\odot }}
\newcommand{\lsun}{{\rm\,L_\odot }}
\newcommand{\ngcs}{NGC~147 }
\newcommand{\ngcf}{NGC~185 }
\DeclarePairedDelimiter\abs{\lvert}{\rvert}%

\title[A genetic approach to NGC~147, NGC~185 and CassII]{NGC~147, NGC~185 and CassII: a genetic approach to orbital properties, star formation and tidal debris}
\author[Arias et al.]{
{\parbox{\textwidth}{
Veronica Arias$^{1,2}$\thanks{E-mail:v.arias@uniandes.edu.co}, 
Magda Guglielmo$^1$,
Nuwanthika Fernando$^1$, 
Geraint F. Lewis$^1$, 
Joss Bland-Hawthorn$^1$,
Nicholas F. Bate$^{1,3}$,
Anthony Conn$^1$,
Mike J. Irwin$^4$,
Annette M. N. Ferguson$^5$, 
Rodrigo A. Ibata$^6$,
Alan W. McConnachie$^7$, 
Nicolas Martin$^{6,8}$ 
\\}}
\vspace{0.1cm}\\
\parbox{\textwidth}{
$^1$Sydney Institute for Astronomy, School of Physics, A28, The University of Sydney, NSW 2006, Australia \\
$^2$Departamento de F\'isica, Universidad de los Andes, Cra. 1 No. 18A-10, Edificio Ip, Bogot\'a, Colombia\\
$^3$  School of Physics, The University of Melbourne, Parkville, VIC 3010, Australia. \\
$^4$Institute of Astronomy, Madingley Road, Cambridge, CB3 0HA, U.K. \\
$^5$Institute for Astronomy, University of Edinburgh, Royal Observatory, Blackford Hill, Edinburgh, EH9 3HJ, U.K. \\
$^6$Observatoire de Strasbourg, 11, rue de l'Universit\'e, F-67000, Strasbourg, France\\
$^7$NRC Herzberg Institute for Astrophysics, 5071 West Saanich Road, Victoria, British Columbia, Canada, V9E 2E7 \\
$^8$Max-Planck-Institut für Astronomie, Königstuhl 17, D-69117 Heidelberg, Germany\\
}}

\date{Accepted xxx. Received xxxx; in original form xxxx}

\pagerange{\pageref{firstpage}--\pageref{lastpage}} \pubyear{2002}

\maketitle

\label{firstpage}

\begin{abstract}
NGC~147, NGC~185 and Cassiopeia~II (CassII) have similar positions in the sky, distances and measured line of sight velocities. This proximity in phase space suggests that these three satellites of M31 form a subgroup within the Local Group.
Nevertheless, the differences in their star formation history and interstellar medium, and the recent discovery of a stellar stream in NGC~147, combined with the lack of tidal features in the other two satellites, are all indications of complex and diverse interactions between M31 and these three satellites.
We use a genetic algorithm to explore the different orbits that these satellites can have and select six sets of orbits that could best explain the observational features of the NGC~147, NGC~185 and CassII satellites.
The parameters of these orbits are then used as a starting point for N-body simulations. We present models for which NGC~147, NGC~185 and CassII are a bound group for a total time of at least one Gyr but still undergo different interactions with M31 and as a result NGC~147 has a clear stellar stream whereas the other two satellites have no significant tidal features.   
This result shows that it is possible to find solutions that reproduce the contrasting properties of the satellites and for which NGC~147-NGC~185-CassII have been gravitationally bound.
\end{abstract}


\section{Introduction}

Satellite galaxies are useful tools for testing the current paradigm of galaxy formation, where large halos are assembled from smaller halos in a sequence of accretion events. Some of these sub-halos are completely destroyed in the course of time, while others may survive and some can be observed in the form of satellite galaxies, orbiting a central host. As a possible witness of past accretion events, the satellite population provides effective constraints on galaxy formation. In particular, the Local Group offers a unique laboratory for understanding the properties of these galaxies and comparing them with the results of N-body simulations.
The current models of structure formation predict that galaxies have inhomogeneous stellar halos, containing not only satellites and globular clusters, but also tidal features resulting from the interactions between sub-halos and their central host. 
Tidal streams are observed in both the Milky Way \citep{1994Natur.370..194I, 2001ApJ...547L.133I, 2006ApJ...642L.137B} and M31 \citep{2008MNRAS.390.1437C,2001Natur.412...49I,Ibata2007}, {\color{black} as well as in isolated galaxies in the Local Volume \citep{2010AJ....140..962M}} and are predicted by simulations \citep{2005ApJ...635..931B}.

Apart from the interactions with their host, there is also evidence for interactions or past mergers between satellites. 
Traces of these past satellite-satellite interactions appear in the Leo IV and Leo V system and in the And II galaxy in the form of stellar streams \citep{2010ApJ...710.1664D,2014Natur.507..335A}, or as a shell structure in the Fornax galaxy \citep{2004AJ....127..832C,2012ApJ...756L...2A}. 
Such evidence suggests that some of these satellites have been members of bound pairs. 
Hierarchical substructures of satellites are expected within the $\Lambda$CDM paradigm, as high resolution N-body simulations show that a significant fraction of satellites are members of bound pairs orbiting their host \citep{2007MNRAS.379.1475S,2013MNRAS.428..573S} and that some satellites might have been accreted as a group \citep{2008MNRAS.385.1365L,2008ApJ...686L..61D}. 
The Magellanic Clouds, for example, are believed to have been the central galaxies of a more extended group which contained other Milky Way satellites, like the Ursa Minor or Draco dwarfs \citep{1976MNRAS.174..695L,1995MNRAS.275..429L,2009IAUS..256..473D}. 
\cite{2006AJ....132..729T} proposed seven `associations of dwarfs' in the Local Group or in the local Hubble flow, like the NGC 3109 subgroup \citep{1999A&ARv...9..273V,2006AJ....132..729T,2013A&A...559L..11B}. 
Such associations present strong spatial and kinematic correlations and appear to have high mass-to-light ratios.  
\cite{2013MNRAS.431L..73F} showed that $\sim30$ per cent of Local Group satellites, brighter than $M_{v}=-8$, are likely in bound pairs. 
This analysis also reveals interesting galaxy pair candidates. 

In addition to the potential association of the Magellanic Clouds, 
confirmed by recent dynamical studies of their interactions \citep{2012MNRAS.421.2109B,2012ApJ...750...36D,2014GUGLIELMO}, some of the M31 satellites are possible members of bound pairs, like And~I/And~III and NGC~147/NGC~185 \citep{2013MNRAS.431L..73F}.  
The suggestion that the later form a bound pair was already proposed by \cite{1998AJ....116.1688V}, who argued that due to the proximity in positions, distances and velocities, these satellites are likely to be associated. 
\ngcs and \ngcf have an angular separation of $1^{\circ}$ ($\sim 11 \kpc $ in projected distance); estimated distances from the Milky Way of $712\pm20$ kpc and $620\pm19$ respectively \citep{2012ApJ...758...11C}; systemic heliocentric velocities equal to $-193.1\pm0.8\kms$ for \ngcs and $-203.9\pm1.1\kms$ for \ngcf  \citep{2013ApJ...768..172C}. 
In addition, \ngcs and \ngcf have similar masses, of $5.6\times10^{8}\msun$ and $7.8\times10^{8}\msun$ respectively \citep{2010ApJ...711..361G} and for most of their lifetime, they have had a similar star formation history \citep{2005AJ....130.2087D}; both are dominated by old stars and contain populations of intermediate aged Asymptotic Giant Branch stars \citep{1990AJ.....99...97S,1990AJ....100..108S,1997AJ....113.1001H,2005AJ....130.2087D}.
Evidence against their bound state comes from the recent star formation history of these galaxies and their radically different Interstellar Medium (ISM). 
While \ngcs shows a distinct lack of gas and no recent star formation (the youngest stars formed about 1 and 3 Gyr ago), \ngcf presents continuous star formation and has a gas rich environment \citep{2014ApJ...789..147W}.

Results based on the Pan-Andromeda Archaeological Survey \citep[PAndAS][]{2009Natur.461...66M} help to unravel the mystery of this pair. 
Designed to map the entire stellar halo of M31 and M33, the PAndAS data affords a detailed characterisation of \ngcs and NGC~185, and led to the discovery of a prominent stellar stream in \ngcs, while \ngcf does not present any tidal distortions (\cite{2014ApJ...780..128I}, Irwin et al., in preparation)
This could suggest that \ngcs experienced a close encounter with M31's disk, while the other satellite did not. 
This scenario might also explain the lack of gas in \ngcs because it could have been stripped away by the main galaxy during the encounter. 
Additional studies on this system found globular clusters in the region between these two satellites \citep{2013MNRAS.435.3654V}, and in a recent paper, \cite{2014MNRAS.445.3862C} provide a deep wide-field analysis of the two dwarf ellipticals (dEs) to study their structures, stellar populations and chemical properties. 
Their work constrains the properties of the dEs down to $\sim3$ mag below the red giant tip and confirms that NGC~147 is tidally disrupted while NGC~185 is not. 

The recently discovered dwarf spheroidal (dSph) galaxy CassII makes this system even more interesting (Irwin et al., in preparation). Unlike CassIII, which lies close in projection to the two dEs but with a much greater velocity \citep{2013ApJ...772...15M,2041-8205-793-1-L14}, CassII presents strong spatial and kinematic correlations with \ngcs and NGC~185: having a distance from the Milky Way of $\sim680$ kpc \citep{2012ApJ...758...11C}, CassII is only 35 kpc away from \ngcs and 63 kpc from NGC~185. 
As suggested by \cite{2013ApJ...768..172C}, the line-of-sight velocity of this newly discovered galaxy ($\sim-139\kms$) is also very close to that of \ngcs and NGC~185, strongly suggesting that the history of this satellite is related to that of the other two. 
 
In the current paper, we explore the possibility that NGC~147, \ngcf and CassII formed a bound group that is now in the process of dissolution, after an encounter with M31. 
This scenario, first proposed by Irwin et al. (in preparation), could explain the observed similarities in NGC~147's and NGC~185's star formation histories, as well as the proximity between these satellites and CassII. 
The encounter with M31, which broke up the group, might be responsible for the formation of the stream in NGC~147. 
As neither \ngcf nor CassII present any evidence of tidal disruption,
with no streams visible down to $32 {\rm\,mag\,arcsec^{-2} }$, which is the surface brightness limit of the data, these two galaxies probably never came as close as \ngcs to M31's disc. 
By translating these considerations into model requirements, we use a genetic algorithm to explore different orbital parameters. 
Unlike the Milky Way satellites, for which the full 3D kinematic information is available to constrain the orbital properties of some of them \citep{2012ApJ...748..149N,2011ApJ...742..110N}, there are currently no proper motion measurements for most Andromeda satellites.
This lack of complete kinematic information makes it difficult to properly constrain the interaction between NGC~147, NGC~185 and CassII with the Andromeda galaxy without resorting to numerical techniques, like the genetic algorithm we use in this work.
In addition, there are also uncertainties in the distances and mass estimates of the satellites.
Therefore we consider their tangential velocities, the distance from the Milky Way and the total mass of NGC~147, NGC~185 and CassII as free parameters that vary within a range given by the observational constraints.

This paper is organised as follows. 
In Section \ref{secNumMod} we introduce the numerical model used to describe both the Milky Way and the M31 potentials. 
In Section \ref{secStatAna}, we determine the likelihood of a chance alignment, confirming that the NGC~147-NGC~185-CassII system is (or was at some point) a bound system. 
We describe in Section \ref{secGA} the genetic algorithm, the different orbital parameters that we consider, and present the resulting orbits. In Section \ref{secNbody} we present the results of N-body simulations that use the orbital parameters selected by the genetic algorithm. 
Finally, in Section \ref{Plane}, we include further constraints on the model to obtain orbits that are co-planar with the plane of satellites of Andromeda \citep{2013ApJ...766..120C,2013Natur.493...62I} and present additional results of N-body simulations. 

\section{Numerical Model}\label{secNumMod}
In this work, we explore the possible orbits of NGC~147, NGC~185 and CassII around the Andromeda galaxy. 
To make this exploration and the integration of the orbits more efficient, we use a rigid potential for both Andromeda and the Milky Way and, in the first part of this work we consider the satellite galaxies as point masses.

The Andromeda potential is described as a three component model, first proposed by \cite{2006MNRAS.366..996G}.  
The dark matter halo is described as a Navarro-Frenk and White (NFW) potential \citep{1997ApJ...490..493N} given by
\begin{equation}
\Phi_{\rm{halo}}(r)=-\frac{\rm{GM_{\rm{halo}}}}{r}\log \left(\frac{r}{r_{\rm{halo}}}+1\right),
\label{NFW}
\end{equation}   
The disk component of the potential is given by
\begin{equation}
\Phi_{\rm{disk}}(r)=-2{\pi}G\Sigma_{0}{r_{\rm{disk}}}^2\left[\frac{1-e^{-r/r_{\rm{disk}}}}{r}\right]
\end{equation}   
and the bulge component follows a Hernquist profile \citep{1990ApJ...356..359H}
\begin{equation}
\Phi_{\rm{bulge}}(r)=-\frac{\rm{GM_{{bulge}}}}{r_{\rm{bulge}}+r}\mbox{.}
\label{Hernquist}
\end{equation}\\

The Milky Way potential is described by a Hernquist bulge, a NFW halo potential (see equations (\ref{Hernquist}) and (\ref{NFW})), 
and a Galactic disk described by a Miyamoto-Nagai potential \citep{1975PASJ...27..533M} given by
\begin{equation}
\Phi_{\rm{MNdisk}}(R,z)=-\frac{\rm{GM_{MNdisk}}}{\left(R^2+\left(r_{\rm{MNdisk}}+{\sqrt{(z^2+b^2)}}\right)^2\right)^{1/2}}
\end{equation}\\

All the parameters we use are listed in Table \ref{tabPot}.
The models described above for M31 and the Milky Way are kept the same in all the different steps of our analysis. 
In the case of the N-body simulation (see Section \ref{secNbody}), the \textsc{Gadget-2} code \citep{2005MNRAS.364.1105S} has been modified in order to include M31 and the Milky Way as rigid potentials. 
In all our models, the Milky Way is kept fixed in its current position. 
To test the validity of this approximation, we ran test simulations with a moving Milky Way and found no
significant difference with simulations with a fixed Milky Way.

Additionally, for all the dynamical analysis, we use a coordinate system centered on M31 and for which M31's disk lies in the $xy$ plane. 
The distances, right ascensions and declinations of the satellites 
are transformed into this coordinate system as described by \cite{2013ApJ...766..120C}.
The measured velocities of the satellites are also transformed into this coordinate system by subtracting M31's heliocentric velocity from the z-component of the satellites' heliocentric velocities \citep{2013ApJ...768..172C}. 

\begin{table}
\centering
\begin{tabular}{l l }
\multicolumn{2}{c}{\textbf{M31}}\\
\hline
\hline
$\rm{M}_{bulge}$   	& $2.86\times{10^{10}}\,\rm{M}_{\odot}$\\
$\rm{r}_{bulge}$	& $0.61\,\rm{kpc}$\\
$\rm{M}_{disk}$		&  $8.4\times{10^{10}}\,\rm{M}_{\odot}$\\
$\rm{r}_{disk}$		& $5.4\,\rm{kpc}$\\
$\Sigma_{0}$		& $4.6\times10^{8}\,\rm{M}_{\odot}\rm{kpc^{-2}}$\\
$\rm{M}_{halo}$		&  $103.7\times{10^{10}}\,\rm{M}_{\odot}$\\
$\rm{r}_{halo}$		& $13.5\,\rm{kpc}$\\
&\\
\multicolumn{2}{c}{\textbf{MW}}\\
\hline
\hline
$\rm{M}_{bulge}$       & $ 3.4\times{10^{10}}\,\rm{M}_{\odot}$   \\
$\rm{r}_{bulge}$       & $ 0.7\,\rm{kpc}$                        \\
$\rm{M}_{MNdisk}$        & $ 10.0\times{10^{10}}\,\rm{M}_{\odot}$  \\
$\rm{r}_{MNdisk}$        & $ 6.65\,\rm{kpc}$                       \\
$\rm{b} $              & $ 0.26\,\rm{kpc}$                       \\
$\rm{M}_{halo}$        & $ 91.36\times{10^{10}}\,\rm{M}_{\odot}$   \\
$\rm{r}_{halo}$        & $ 24.54\,\rm{kpc}$                        \\
$\rm{d}_{\rm{M31-MW}}$ & $ 779\,\rm{kpc}$                        \\   
\hline
\hline
\end{tabular}
\caption{Parameters used for the M31 and Milky Way potential. The M31 parameters are consistent with \protect\cite{2006MNRAS.366..996G} and the Milky Way disk and bulge parameters are taken from \protect\cite{2005ApJ...635..931B} and are consistent with the parameters presented by \protect\cite{0004-637X-794-1-59}.}
\label{tabPot}
\end{table}

\section{Statistical analysis}
\label{secStatAna}
NGC~147, NGC~185 and CassII have similar positions and line of sight velocities
{\color{black} \footnote[1]{{\color{black}For this study, we have used the best distance estimate to CassII from \cite{2012ApJ...758...11C}(681+32-78 kpc). We note that Irwin et al (in preparation) have revised this distance using the g-band luminosity function to 575+-30 kpc, corresponding to the lower, second peak in the distance probability distribution presented by \cite{2012ApJ...758...11C}. 
As CassII is a minor player in the dynamical interactions studied in this paper, we employ the earlier distance as it will not influence the conclusions of the paper.}} 
\citep{2012ApJ...758...11C,2013ApJ...768..172C}}.
This proximity in phase space suggests that these three satellites are (or were at some point in their history) a bound group.
Nevertheless, due to uncertainties in both their masses and proper motions, we cannot discern between the system being a group versus it being a chance alignment based solely on the observational data.       
Therefore, we use numerical simulations to determine the likelihood of these two possible scenarios.
 
We construct different sets of orbits by randomly selecting values for the unknown proper motions within a given velocity range, then integrating the orbits backwards in time for 8 Gyr, and finally randomly rotating the velocity vector. 
The orbits are then integrated forwards, and the distances between the three satellites are computed for each time step. 
We finally sum the time intervals when all the inter-satellite distances are smaller than a threshold value, to obtain the percentage of time that the satellites could be `observed' as being in a group. 
This process is repeated for 10000 sets of orbits and the total probability is calculated by dividing the sum of the individual probability for each set of orbits by the total number of orbit sets.
 
For a satellite mass $M = M_*$ (where $M_*$ is the stellar mass of each satellite), a velocity range of $150 \kms$, and for a threshold value of 100~kpc, which is around the distance between NGC~147 and NGC~185, we find that the probability of the satellites having a spatial separation smaller than the threshold is {\color{black}$0.15$ per cent}. 
This value is an upper limit for the `chance alignment scenario' since we count also the cases where the random velocities we chose result in orbits where the satellites form a bound system. 
We repeat the numerical experiment for a satellite mass $M = 10 \times M_*$ and
find that there is no significant variation compared to the probability obtained for $M = M_*$. 
These two results show that it is very unlikely that the current location of the three satellites is a chance alignment.
Additionally, we know that the line of sight velocities of the three satellites are similar. If we include in this analysis a velocity threshold of $100 \kms$, we find that the probability of the current configuration being a coincidence decreases even further to less than $1.0 \times 10 ^{-5}$.

The previous analysis is valid when we assume that the dwarf ellipticals NGC~147 and NGC~185 are a special pair among all the other M31 dwarf spheroidal satellites. 
If on the contrary we consider NGC~147 and NGC~185 as two standard satellites of the M31 population, then, to study the likelihood of having the NGC~147-NGC~185-CassII group, we need to analyse the whole satellite population. 
In order to do so, we repeat the previous analysis for thirty satellites, and look for groups of three or more satellites. 
We use again a distance threshold value of 100~kpc and a velocity threshold value of $100 \kms$. 
We construct 50000 sets of orbits and find that the probability of at least three of the satellites having a spatial separation smaller than the threshold and a velocity difference smaller than $100 \kms$ is {\color{black} $0.37$ per cent}. 
We can therefore conclude that it is highly unlikely that the Andromeda satellites NGC~147, NGC~185 and CassII are a chance alignment, and it is therefore reasonable to assume that they are (or were at some point in their history) a bound group. 
This information can be used to constrain the orbits of these three satellites in the Andromeda potential.

\section{Exploring the Parameter Space}
\label{secGA}

\begin{figure}
\begin{center}
\includegraphics[width=0.5\textwidth]{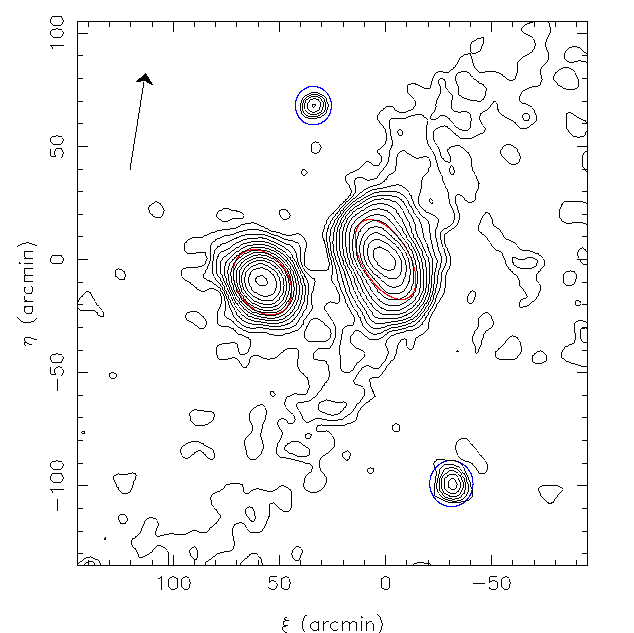}
\caption{The surface density variation of RGB stars
in a 4 $\times$ 4 degree region around NGC~147 and NGC~185. The nominal
tidal radii of both are indicated with the red ellipses, where all parameters
were taken from the compilation of \protect\cite{1998ARA&A..36..435M}. The blue-circled
object to the bottom right is the recently discovered dwarf spheroidal
And~XXV \protect\citep{2011ApJ...732...76R} while the blue-circled object to the top
is the newly discovered dwarf spheroidal CassII. Contours start at 0.25
arcmin$^{-2}$ above background (0.3 arcmin$^{-2}$) and thereafter are incremented
nonlinearly in steps {\color{black} of $0.15 \times 1.5^{i-1}$ arcmin$^{-2}$} to avoid
excessively crowding the inner contours. {\color{black}There are 1,134,386 RGB candidate stars in the image.
Note that the plane of satellite galaxies \citep{2013ApJ...766..120C,2013Natur.493...62I} runs north-south in this image ({\color{black} as indicated by the arrow)}.}
Figure from {\color{black}Irwin et al. (in preparation).}
}
\label{fig_cont}
\end{center}
\end{figure}

In this work, a genetic algorithm \citep{Holland:1975,1995ApJS..101..309C} combined with a simple point mass integration, is used to identify the orbit of the three satellite galaxies. The basic idea of the genetic algorithm (GA) is to emulate the biological concept of evolution, by studying the evolution of an initial set of solutions (\emph{individuals}). 
The strongest individuals are identified according to their ability to satisfy the model requirements. 
This is expressed in terms of a merit function (or ``fitness function'') that is appropriately chosen to describe a particular model. 

In the particular case of NGC~147-NGC~185-CassII, the free parameters are their tangential velocities, their distances from the Milky Way and their masses.

\begin{itemize}
\item[-] \emph{Tangential velocity}: These are chosen in the plane perpendicular to the line of sight velocity, using the equation 
\begin{equation}
V_{{\rm{los}}_{{\rm{x}}}}v_{\rm{x}}+V_{{\rm{los}}_{{\rm{y}}}}v_{\rm{y}}+V_{{\rm{los}}_{{\rm{z}}}}v_{\rm{z}}=0  ,
\label{eqPlane}
\end{equation}
where $(V_{{\rm{los}}_{{\rm{x}}}},V_{{\rm{los}}_{{\rm{y}}}},V_{{\rm{los}}_{{\rm{z}}}})$ are the line-of-sight velocity components and $(v_{\rm{x}},v_{\rm{y}},v_{\rm{z}})$ are the unknown components of the tangential velocity. 
In particular, the $v_{\rm{x}}$ and $v_{\rm{y}}$ components for each satellite are randomly selected between the range described in Table \ref{tab:ParRange}, while the $v_{\rm{z}}$ component is directly calculated from equation \ref{eqPlane}.
{\color{black} It is important to note that we select the unknown components of the tangential velocity in a coordinate system that is centred in the Milky Way and for which the z-axis corresponds to the line that connects M31 and the Milky Way. 
In this coordinate system, the xy plane is parallel to the tangential plane for M31, and therefore, the unknown $v_{\rm{x}}$ and $v_{\rm{y}}$ tangential velocity components are the dominant components, and, due to the conditions imposed to calculate the $v_{\rm{z}}$ component, it will not significantly affect the magnitude range of the total tangential velocity of the satellites.}\\ 
\item[-] \emph{Distance from the Milky Way}: The distances of NGC~147, NGC~185 and CassII from the Milky Way have been tightly constrained by \cite{2012ApJ...758...11C}. 
These measurements strongly affect the relative position in the M31 reference frame and hence, the present day position of each satellite galaxy with respect to its host. 
In order to investigate the influence of the distances and their uncertainties on the orbit, the GA considers these distances as free parameters that vary within the observational error range.\\
\item[-] \emph{Total Mass}: The last set of parameters consists of the total mass of the three satellites. 
We have good estimates for NGC~147 and NGC~185's stellar mass only within the 2$R_e$ \citep{2010ApJ...711..361G}, values that can be an underestimation of their stellar content. 
In addition, although these two satellites are not expected to be dark matter dominated within their luminous radius, it is reasonable to assume that there is a dark matter contribution to their total masses.
These are fundamental parameters for orbital calculation and different mass models need to be explored in order to constrain the past evolution of these satellites. 
Therefore, the mass of NGC~147 and NGC~185 are allowed to vary between a minimum value given by their stellar mass $M_{NGC 147}=0.56\times10^{9}\,\rm{M_{\odot}}$ and $M_{NGC 185}=0.72\times10^{9}\,\rm{M_{\odot}}$, from \cite{2010ApJ...711..361G}. The maximum is instead assumed to be two orders of magnitude greater than the stellar component. 
Recent estimates by \cite{2014ApJ...783....7C}, put the mass of CassII at $0.02\times{10^9}\,{\rm{M_\odot}}$ with a lower limit of $0.009\times{10^9}\,{\rm{M_\odot}}$. Therefore, in our model, we allow its total mass to vary between $0.01-1.0\times{10^9}\,{\rm{M_\odot}}$, where the lower limit is consistent with the observational error, while the upper limit accounts for the dark matter component.
\end{itemize}

\begin{table}
\begin{tabular}{c c c c c}
\hline
&$v_{\rm{x}}$ & $v_{\rm{y}}$& D$_{MW}$&${\rm{M_{Tot}}}$\\
&(${\rm{km\,s}^{-1}}$) &(${\rm{km\,s}^{-1}}$)& (kpc) & ($10^{9}$ M$_{\odot}$)\\
\hline
NGC147 & $[-200,\,200]$ & $[-200,\,200]$& $[693,\,733]$& $[0.56,\,9.0]$\\
NGC185 & $[-200,\,200]$ & $[-200,\,200]$& $[602,\,639]$& $[0.72,\,9.0]$\\
CassII & $[-200,\,200]$ & $[-200,\,200]$& $[603,\,713]$& $[0.01,\,1.0]$\\
\hline
\end{tabular}
\caption{Parameter range used in the genetic algorithm.}
\label{tab:ParRange}
\end{table}
 
\subsection{The Selection of the Orbits} \label{subsecGAparam}
 As discussed in Section \ref{secStatAna}, the possibility of a chance alignment is very unlikely and hence, the proximity in distance and line-of-sight velocity could be an indication that these three satellites are or have been bound to each other. By only considering NGC~147 and NGC~185, \cite{1998AJ....116.1688V} firstly explored this possibility, concluding that the mass, morphology and velocity of these galaxies strongly suggest that they form a bound pair. 
This initial hypothesis is corroborated by the recent results presented by \cite{2013MNRAS.431L..73F}, who identified NGC~147 and NGC~185 as likely bound galaxies. 
Based on these works, we use the genetic algorithm to select orbits such that NGC~147 and NGC~185 are bound at least once in the past 8 Gyr.
 
In addition, the recent discovery of CassII, with similar position and velocity to those of NGC~147 and NGC~185, opens the possibility that this satellite galaxy is another member of the group and is therefore bound to one of the two dEs or to their centre of mass \citep{2013ApJ...768..172C,2013MNRAS.430..971W}. 
{\color{black} If a system is gravitationally bound, then the total energy is
negative, and we can use this as a condition on the orbits of the
satellites. We can be certain that the system is bound if the ratio $2K/\abs U<<1$, where $\rm{K}$ and $\rm{U}$
are the kinetic and potential energy respectively, whereas $2K/\abs
U>>1$ implies that the system is not bound. With these conditions in
mind }
the fitness function is given by the sum over all time-steps of the ratio $|U|/(|U|+2K)$ averaged for the total integration time. This ratio will tend to one when $2K\ll|U|$ and zero otherwise. The final expression for this component of the fitness function is :
\begin{equation}
f_{b}=\frac {1}{T}\sum _t ^T{\left(\frac{\abs U}{\abs{U}+2K}\right)_t}
\label{eqFitnessBound}
\end{equation}
where T=8 Gyr. 
{\color{black} This binding fraction tends to unity when the system is fully bound
for $100\%$ of the integration time, and smaller values represent less
bound, or only temporarily bound orbits
}

In addition, the existence of the stellar stream in NGC~147 (\cite{2014ApJ...780..128I}, Irwin et al., in preparation) allows for further considerations on the orbit of these galaxies. 
In that paper, it is reasoned that, if the formation of the stream was due only to the interaction between NGC~147 and NGC~185, then both of them should appear equally disrupted, since they have a similar mass. 
As this is not the case, the formation of the stream could be the result of a close encounter between NGC~147 and M31's disk, while the lack of tidal disruption in NGC~185 and in CassII suggests that a close encounter between these two and their host galaxy is less likely. 
It may be argued that this assumption naturally disproves the hypothesis that these two galaxies are bound today, because any past encounter with M31 would probably disrupt the binary pair \citep{1998AJ....116.1688V}. 
However, it is still possible that this group is now in, or has just completed, the process of dissolution and therefore, the condition given by equation (\ref{eqFitnessBound}) might still be satisfied for some time in the past. 

These hypotheses on the formation of the stream have been translated into further conditions on the orbits. 
In particular the requirements are on the distances of the satellites from their host at the time of the encounter. 
The latter can be constrained using information on the star formation history of the NGC~147/NGC~185 system.  
While young stars \citep[$\sim400$ Myr][]{1998AJ....115.1462M} have been observed in NGC~185, the most recent star formation activity in NGC~147 occurred between 1 and 3 Gyr ago \citep{2005AJ....130.2087D,1997AJ....113.1001H}, probably after an encounter with M31. 
During this encounter, NGC~147 needs to get very close to the M31 disk, so that this interaction can strip its gas away and, can eventually, lead to the formation of the stream. 
Due to the uncertainties of the M31 disk parameters, the GA looks for orbits where NGC~147 has at least one encounter with M31, getting closer than 60kpc from its center but never closer than 20 kpc (which is a rough estimate of the disk radius). 
Additionally, by considering the age of the young stars, it is reasonable to assume that one or more encounters with M31 occurred between 3 and 1 Gyr ago. 
These requirements are expressed in the following equation: 

\begin{equation}
f_{NGC147}=\frac{1.0}{1.0+\left(\frac{(d_{NGC147}-\bar{d_1})}{\sigma_{d_1}}\right)^2},
\label{eqFitnessNGC147}
\end{equation}

where $d_{NGC147}$ is the NGC~147 distance {\color{black} to M31} at the time of the encounter in the time range of $-3.0$ and $-1.5$ Gyr; and where {\color{black}$\bar{d_1}=40$ kpc and $\sigma_{d_1}=20$ kpc} are constants chosen to constrain $d_{NGC147}$ to the distance range {\color{black} 20 kpc $< d_{NGC147} <$ 60 kpc} described above. 

There are additional requirements on the distances from NGC~185 and CassII to M31's centre. As these satellites do not present evidence of any tidal disruption, their distances to M31 at the time of the NGC~147/M31 encounter must have been greater than $d_{NGC147}$.
We therefore, use the following two conditions
\begin{equation}
f_{NGC185}=\frac{1.0}{1.0+\left(\frac{(d_{NGC185}-d_2)}{{40\,{kpc}}}\right)^2}
\label{eqFitnessNGC185}
\end{equation}
\begin{equation}
f_{CassII}=\frac{1.0}{1.0+\left(\frac{(d_{CassII}-d_2)}{{40\,{kpc}}}\right)^2}
\label{eqFitnessCassII}
\end{equation}
{\color{black} where $\rm{d_{NGC185}}$ and $\rm{d_{CassII}}$ are respectively 
NGC~185 and CassII distances from M31's center.
The term $\rm{d_2}=\rm{d_{NGC147}}+70.0\,{kpc}$ favors orbits for which neither NGC~185 nor CassII get closer than NGC~147 to the M31 centre.}

{\color{black} In conclusion, to construct the merit function used by the genetic algorithm, we combine three optimization conditions: the time of the encounter between the satellites and M31, their distance during this encounter and the fact that the three satellites must have been a bound group at least for some time (as described above). Therefore the genetic algorithm selects orbits that better satisfy all the requirements, by looking at the solutions that have the highest values of the merit function defined as
\begin{equation}
F={f_b}*{f_{NGC147}}*{f_{NGC185}}*{f_{CassII}}.
\label{eqFitnessTot}
\end{equation}
}
\begin{figure*}
\includegraphics[width=14cm]{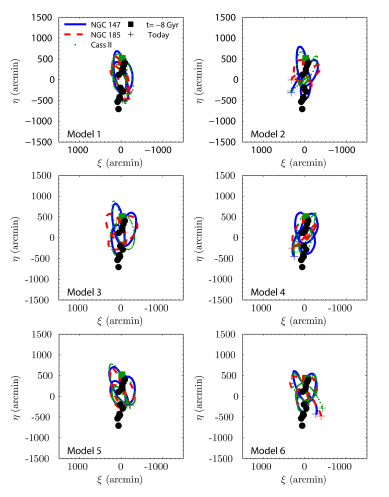}
 \caption{Orbits of NGC~147 (blue sold line), NGC~185 (red dashed line) and CassII (green dotted line) projected on the tangential plane. The black circles indicate the positions of other satellites belonging to the M31 co-rotating plane of satellite galaxies. {\color{black}The squares and crosses represent the initial and final positions of NGC~147 (blue), NGC~185 (red) and CassII (green).}}
\label{figOrbits}
\end{figure*}

\begin{figure*}
\includegraphics[width=14cm]{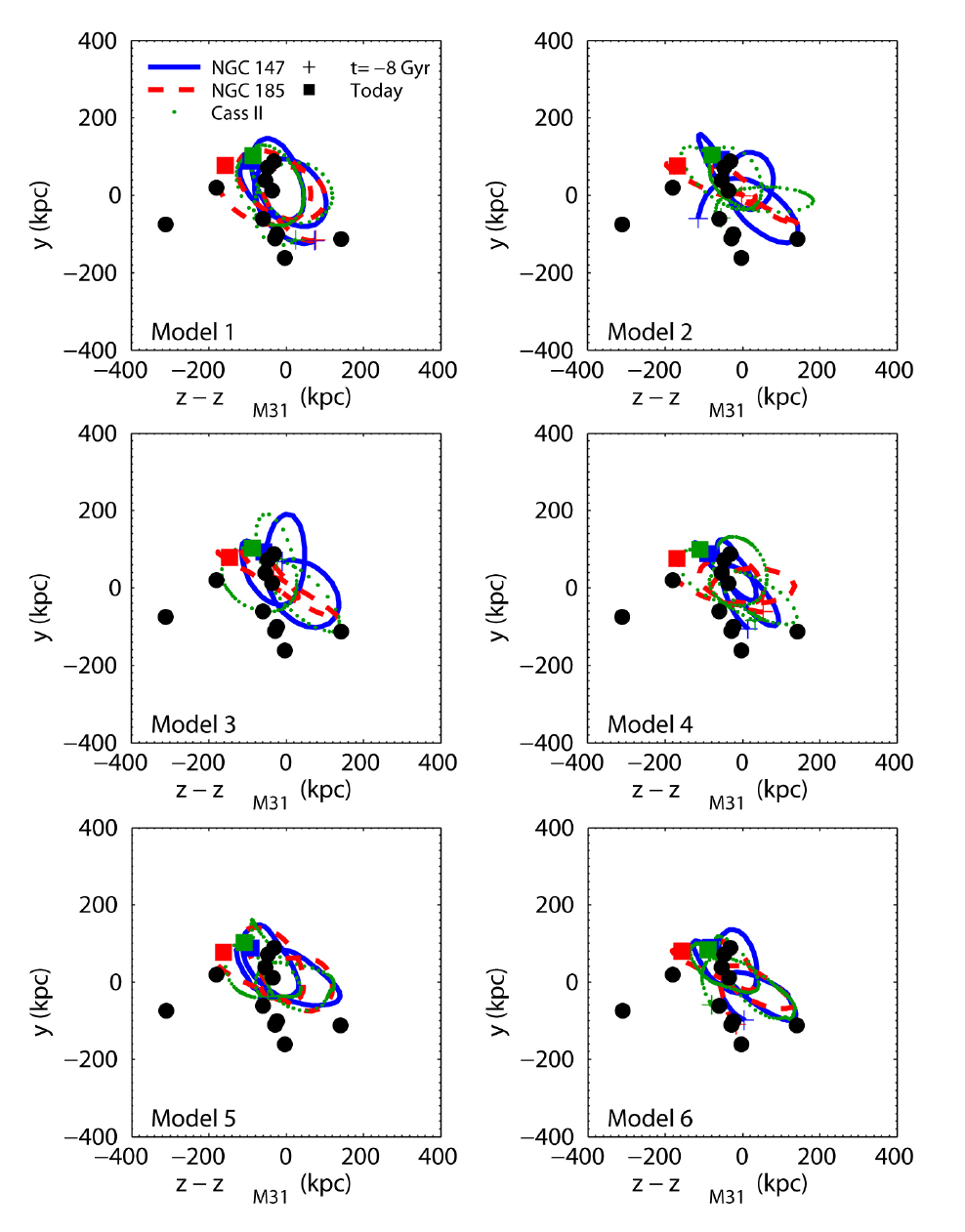}
 \caption{Orbits of NGC~147 (blue sold line), NGC~185
 (red dashed line) and CassII (green dotted line) projected on the yz
 plane. The black circles indicate the positions of other satellites
 belonging to the M31 co-rotating plane of satellite galaxies, which
 is almost face-on in this projection. The squares and crosses
 represent the initial and final positions of NGC~147 (blue), NGC~185
 (red) and CassII (green).}
\label{figOrbitsFaceOn}
\end{figure*}

\begin{figure*}
\includegraphics[width=14cm]{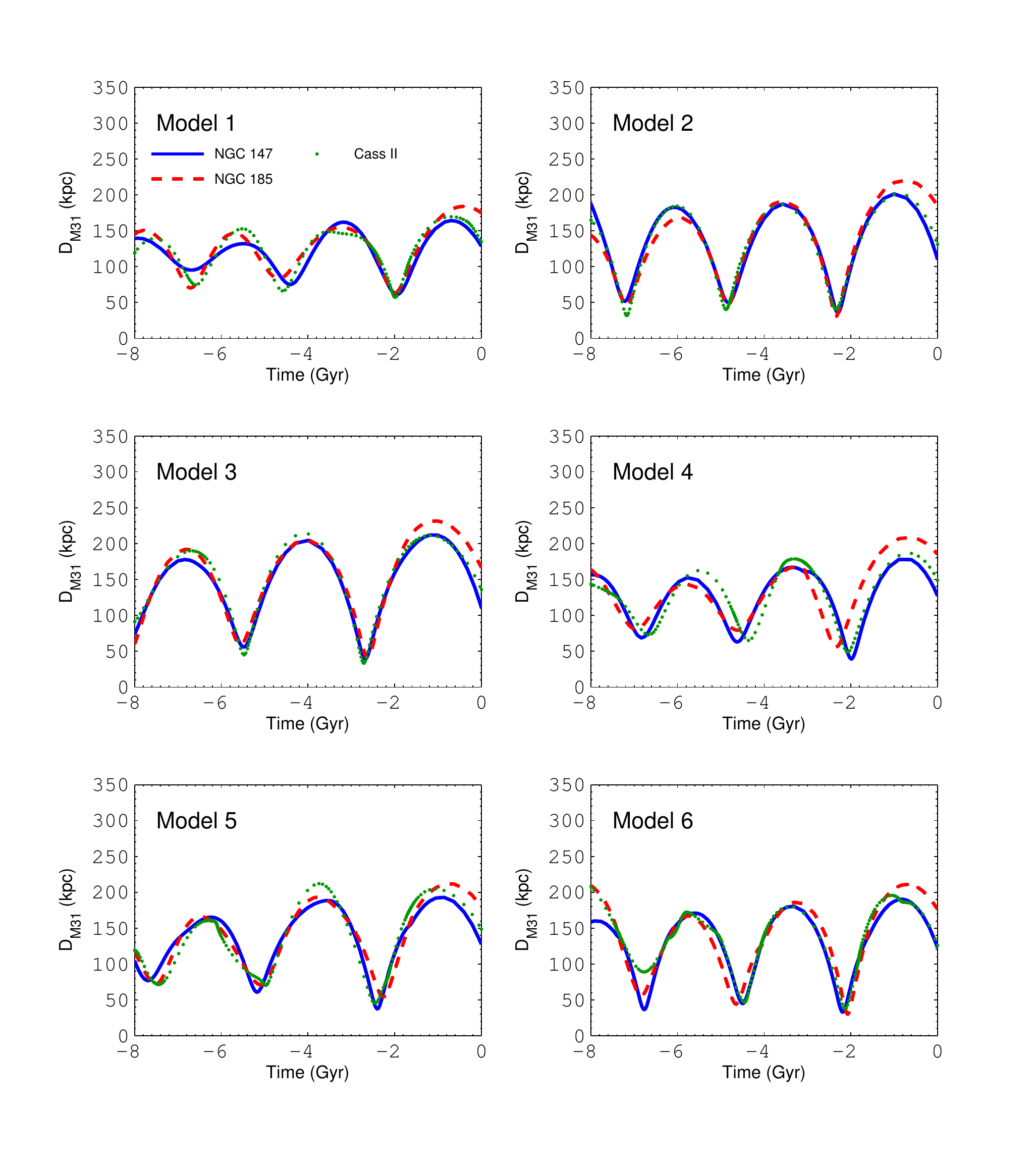}
 \caption{Distance from M31 to NGC~147, NGC~185 and CassII in the
 period of 8 Gyr for each model.}
\label{figDistM31}
\end{figure*}

\begin{table}
\centering
\begin{tabular}{cccccc}
&Galaxy& $\rm{D_{MW}}$ (kpc)& Mass ($10^{10}{\rm{M_{\odot}}}$)& V$_{\rm{t}}$ ($\kms$) \\
\hline
\hline
Model 1 &NGC147 & 693 & 0.88 & 160\\
        &NGC185 & 627 & 0.40 & 146\\
        &CassII & 701 & 0.09 & 199\\
\hline
Model 2 &NGC147 & 730 & 1.02 & 191\\
        &NGC185 & 615 & 0.98 & 108\\
        &CassII & 708 & 0.04 & 120\\
\hline
Model 3 &NGC147 & 728 & 0.91 & 207\\
        &NGC185 & 637 & 0.66 & 155\\
        &CassII & 699 & 0.04 & 140\\
\hline
Model 4 &NGC147 & 695 & 1.01 & 153\\
        &NGC185 & 615 & 0.58 & 137\\
        &CassII & 676 & 0.09 & 88\\
\hline
Model 5 &NGC147 & 694 & 1.00 & 173\\
        &NGC185 & 623 & 1.02 & 128\\
        &CassII & 682 & 0.1 & 154\\
\hline
Model 6 &NGC147 & 698 & 1.03 & 153\\
        &NGC185 & 631 & 0.71 & 114\\
        &CassII & 697 & 0.07 & 149\\
\hline
\hline
\end{tabular}
\caption{Results for the best parameters found by the GA for each model in Figures \ref{figOrbits}, \ref{figOrbitsFaceOn} and \ref{figDistM31} as labeled in Column 1. For each satellite columns 3-5 show the distance from the Milky Way, the total mass and the tangential velocity.}
\label{tabGAmodels}
\end{table}

\begin{table}
\centering
\begin{tabular}{cccccc}
Satellite&  Mass ($10^{10}{\rm{M_{\odot}}}$)& V$_{\rm{t}}$ ($\kms$) \\
\hline
\hline
NGC147 & 0.96 $\pm$ 0.07 & 166 $\pm$ 20\\
NGC185 & 0.77 $\pm$ 0.21 & 121 $\pm$ 19\\
CassII & 0.07 $\pm$ 0.03 & 134 $\pm$ 33\\
\hline
\hline
\end{tabular}
\caption{Results for the error estimates of the parameters found by
the GA. Column 2 and 3 shows the total mass and  the tangential
velocity respectively.}
\label{ParamErrors}
\end{table}

The genetic algorithm was run several times, using 300 individuals and 300 generations, to obtain orbits that best satisfy the conditions described by equations (\ref{eqFitnessBound}) to (\ref{eqFitnessTot}). 
In our case, the free parameters are the masses of the satellites, their proper motions and their exact distance to M31. We vary the possible masses of the satellites from their stellar mass $M = M_*$ to ten times this value. 
For the proper motions of the satellites, we construct velocity vectors that are perpendicular to the measured line of sight velocity and that have magnitudes between $0 \kms$ and $150 \kms$. 
Finally, the initial positions of the satellites in our orbit integrations are chosen within the error of the positions measured by \cite{2012ApJ...758...11C} to account for the observational uncertainties in the distance measurements.     

We chose the six models that best reproduced the orbital constraints.
The masses, tangential velocities, and distances are listed in Table \ref{tabGAmodels} for each satellite in each of the models. 
The genetic algorithm does not give information on the final distribution of the parameters, so we estimate the errors by drawing values of the distances from a Gaussian distribution centered on the observed distances, and running the genetic algorithm for these to obtain a distribution for the parameters which, along with corresponding errors, are listed in Table \ref{ParamErrors}.

Projections in the tangential plane of the orbits corresponding to the six best models are plotted in Figure \ref{figOrbits}  {\color{black} (where the plane of satellites is almost edge on). Additionally, Figure \ref{figOrbitsFaceOn} shows their projections in the YZ-plane (where the plane of satellites is almost face-on)}.
{\color{black} Figure \ref{figDistM31} shows the distance of NGC~147, NGC~185, and CassII from the Andromeda galaxy as function of time. 
In all the six models the satellites have an encounter with M31 between 2 and 3 Gyr ago. 
For models 4 and 5, at the time of these encounters the distance between NGC~147 and M31 is smaller than the distance between NGC~185 and M31, which satisfies the requirement of NGC 147 being in the inner orbit.
However, this does not hold true for all the models. This is because $f_{NGC185}$ and $f_{CassII}$ (equations \ref{eqFitnessNGC185} and \ref{eqFitnessCassII}) 
are not the major constraints in our selection criteria, as the  solutions are those that optimally satisfy all conditions (see equation \ref{eqFitnessTot}). 
Our requirement on the orbits of NGC 185 and Casss II is that during the encounter they lie at a distance where they should not be disrupted by M31.
To test if this indeed happens, we must replace our point mass approximation for the satellites with more realistic N-body simulations.}

\begin{figure}
 \includegraphics{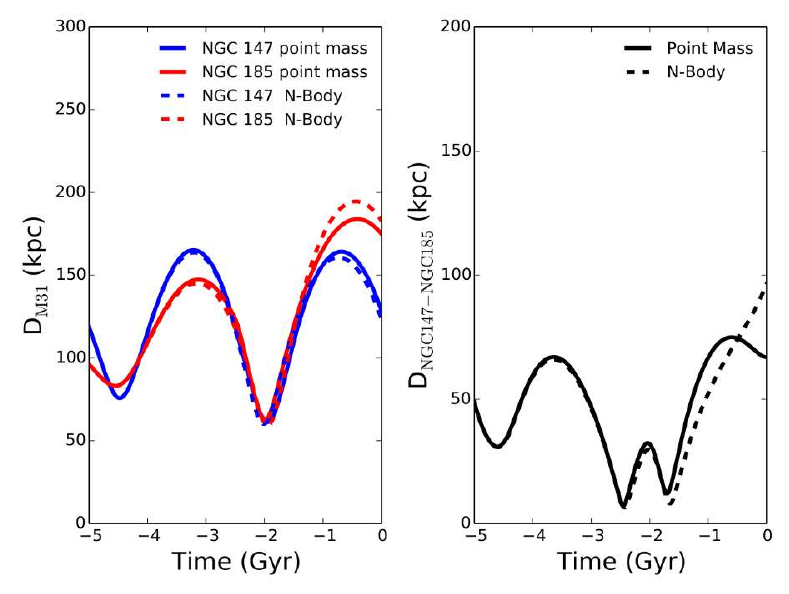}
 \caption{Comparison between the distances obtained with the N-body simulations and the point-mass-approximation model. The left panel shows the distances from M31 to NGC~147 and NGC~185 for the N-body simulation (solid lines) and the point-mass-approximation model (dashed lines). The right panel shows the distance between NGC~147 and NGC~185 for the two models.}
\label{figDistM31Comparison}
\end{figure}

\begin{figure*}
 \includegraphics{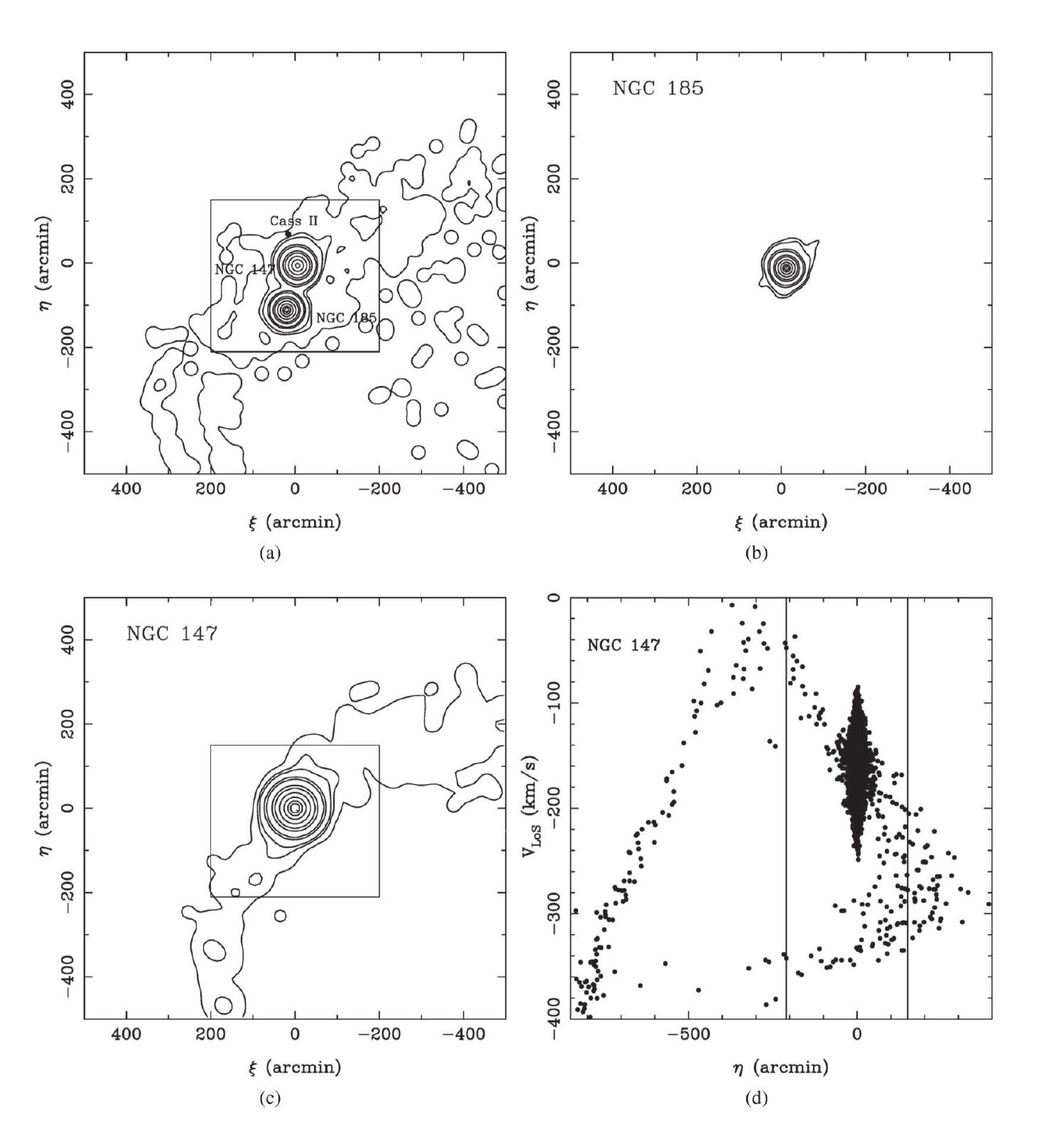}
 \caption{ Mass contour plots of the N-body simulation show the
 projection of NGC~147 and NGC~185 in the tangential plane. Panel (a)
 shows the projection of all the three galaxies, where the solid black
 circle indicates the position of CassII, modelled as point mass.
 Panels (b) and (c) show the individual contributions of NGC~147 and
 NGC~185 to the stream respectively. The black box in panels (a) and
 (b) indicates the region corresponding to the observed stream shown
 in \ref{fig_cont}. Panel (d) shows the line of sight velocity of all
 the NGC~147 particles as a function of $\eta$ coordinate.}
 \label{figContourPlots}
\end{figure*}

\section{N-body simulations}\label{secNbody}

To test if the orbits presented in Figure \ref{figDistM31} result in the NGC~147 observed tidal features and the lack thereof in NGC 185, we study the dynamical interaction of NGC~147 and NGC~185 as N-Body systems. 
Considering its low mass compared with the two dEs, CassII does not play an important role in the interaction with NGC~147 and NGC~185, and therefore it can be approximated as a point mass. 
Andromeda and the Milky Way are included in \textsc{Gadget-2} as fixed potentials (as described in Section \ref{secNumMod}).   
Using GalactICs \citep{2008ApJ...679.1239W}, NGC~147 and NGC~185 are modelled as a truncated dark matter halo and a stellar Sersic profile from the recent estimations described by \cite{2014MNRAS.445.3862C}.
The total mass, the distance from M31 and the present day total tangential velocity of each satellite are given in Table \ref{tabGAmodels} for each of the six models.

For each set of orbits, we perform N-Body simulations using \textsc{Gadget-2} \citep{2005MNRAS.364.1105S}. Since we are interested in characterising the formation of the stream in NGC 147 while NGC~185 suffers no disruption,
we concentrate our analysis in the time interval where the two dEs are bound to each other: 
This helps to put constraints on the mutual interaction between the two, since, if their interaction is too strong, then both will present evidence of a tidal tail or bridge. In all the orbits we analyse, NGC~147 and NGC~185 form a bound pair for the last 5-6 Gyr, so we fix the time interval for all the N-Body simulations to be 5 Gyr.

Additionally, the N-body simulations provide more information on the orbit of these satellites. An important difference between the point mass approximation used in the genetic algorithm and the N-body simulations performed in this section, is that the latter can account for effects due to the changes of the internal structure of the satellites, such as tidal mass loss or merger events. 
These interaction effects result in orbits that differ from the point mass approximation. 
An example of these differences can be seen in the left panel of Figure \ref{figDistM31Comparison}, where the distances between the satellites and M31 are compared for the N-body simulations and the point mass approximation. 
This figure corresponds to the simulation based on the parameters from Model 1 (see Table \ref{tabGAmodels}) and shows that, for the N-body simulation, the final distance between NGC~185 and NGC~147 
is around 30 kpc greater than the point-mass-approximation distance. 
The differences between the solid and dashed lines in the right panel of Figure \ref{figDistM31Comparison} become more evident after the first satellite encounter around 2.5 Gyr ago,
when both NGC~147 and NGC~185 experienced a very close encounter, where the distance between the two was just a few kpc. 
It is likely that this encounter is responsible for the distance
difference between the N-body and the point-mass-approximation
results. 

Despite the 30 kpc difference in the final NGC~147-NGC~185 distance, the orbits obtained with the N-body simulations do not differ in essence from the ones obtained with the point-mass-approximation. 
All the orbital criteria required by the GA are still satisfied. 
We can therefore use the N-body simulation mass distribution to study the satellites' structures. 
To do so, we project the 3D positions of the satellite particles in the tangential plane, to get a 2D distribution that we can compare morphologically with the observations (see Figure \ref{fig_cont}).
The contour plot in panel a) of Figure \ref{figContourPlots} shows the final distribution of the N-Body simulation particles in the NGC~147, NGC~185 projected in the tangential plane. 
A stream of particles, similar to the observed tidal tails, is clearly
visible, extending from the lower-left to the upper-right corner of
the plot, and with a total mass of $ 5.7 \times{10^7}\msun$. This
simulated stream is more extended than the observed one (see Figure
\ref{fig_cont}), and the black box in
Figure \ref{figContourPlots} highlights the region that corresponds to
the observations. 
In panels b) and c) of Figure \ref{figContourPlots}, the contribution of each satellite is plotted; the strongest component of the stream is part of NGC 147 and NGC 185 displays no significant tidal structures. 
In panel d) we plot line of sight velocities of the N-Body simulation particles as predicted by our models. 
The final mass of the \ngcs stellar component is $6.1\times{10^8}\msun$, while \ngcf has a stellar mass of $6.6\times{10^8}\msun$, in agreement with the values quoted in 
\cite{2010ApJ...711..361G} ($\rm{M_{NGC147}}=5.6\times{10^8}\msun$ and $\rm{M_{NGC185}}=7.2\times{10^8}\msun$).
Due to the tidal interaction, \ngcs loses 0.2$\%$ of its
mass within 2$R_e$. \cite{2015ApJ...798...77H} found that \ngcs has an
anomalously high metallicity for its size, suggesting that this system
suffered significant mass loss. Nevertheless, these results are at odds
with new photometric metallicity measures \citep{2014MNRAS.445.3862C}
which show that \ngcs has a lower metallicity, consistent with the
metallicity measured by \cite{2010ApJ...711..361G}.
In addition to that, \cite{2014MNRAS.445.3862C} results on the stellar distribution, density
profile and flat metallicity distribution suggest that \ngcs has
suffered a significant redistribution of material compared to \ngcf. 
The results of our simulations are in agreement with these
observational results.

\section{NGC~147, NGC~185 and CassII and the Andromeda plane of satellites}
\label{Plane}
\begin{figure}
\centering
\includegraphics{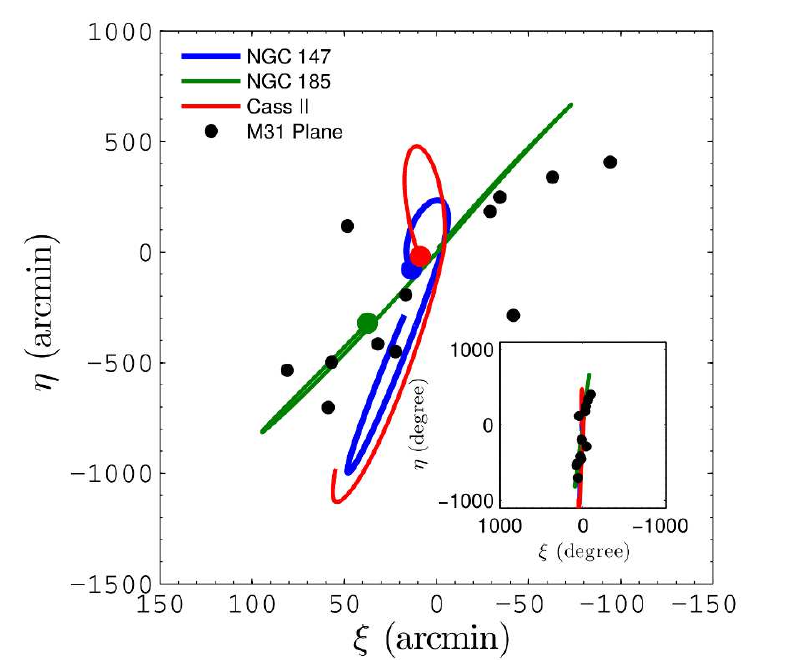}
\caption{Example of orbital models projected on the Sky, when the
tangential velocity components of the three satellites are chosen so
that their present day total velocities lie on the Andromeda plane of
satellites (black points). Note the different scale in the x and y
axis. The small box on the bottom right shows the orbits in the same
axis scale used in Figure \ref{figOrbits}.}
\label{fig:PlaneOrbit}
\end{figure}

\begin{figure*}
\centering
\includegraphics[width=18cm]{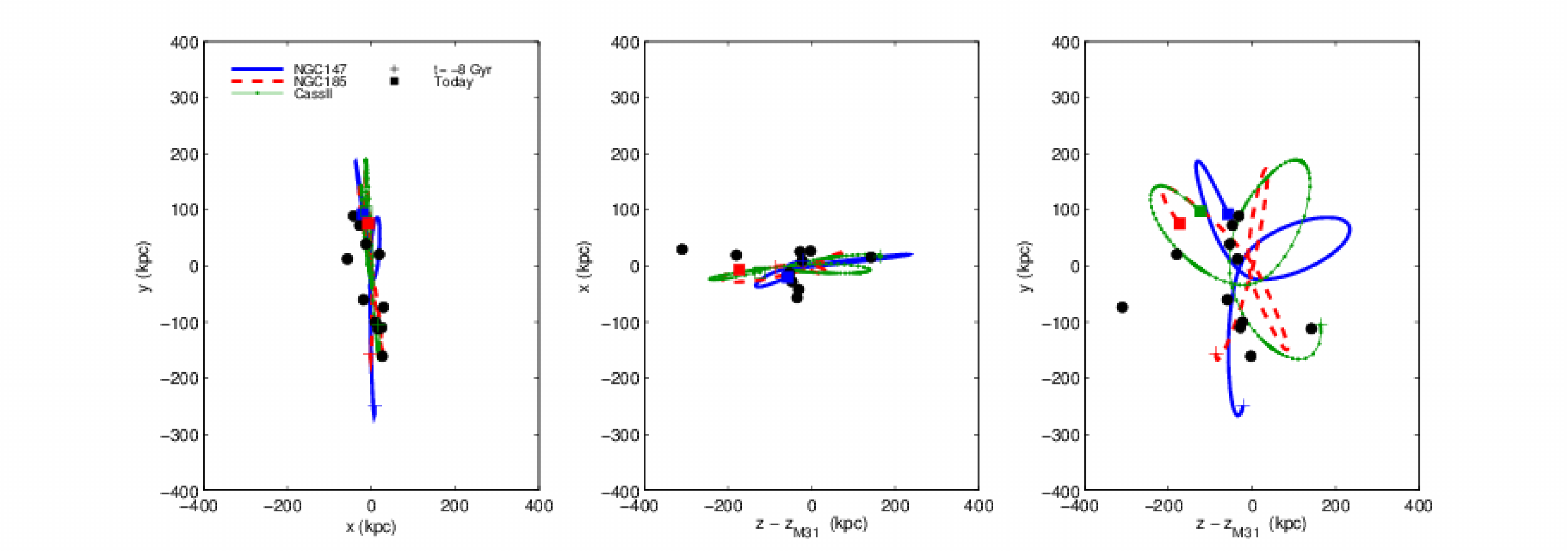}
\caption{{\color{black} Plot of the orbits projected in the XY (left
panel), XZ (middle panel) and YZ (right panel) when the tangential
velocity components of the three satellites are chosen so that their
present day total velocities lie on the Andromeda plane of satellites
(black points).}}
\label{fig:PlaneOrbitXYZ}
\end{figure*}

\begin{figure}
\centering
\includegraphics{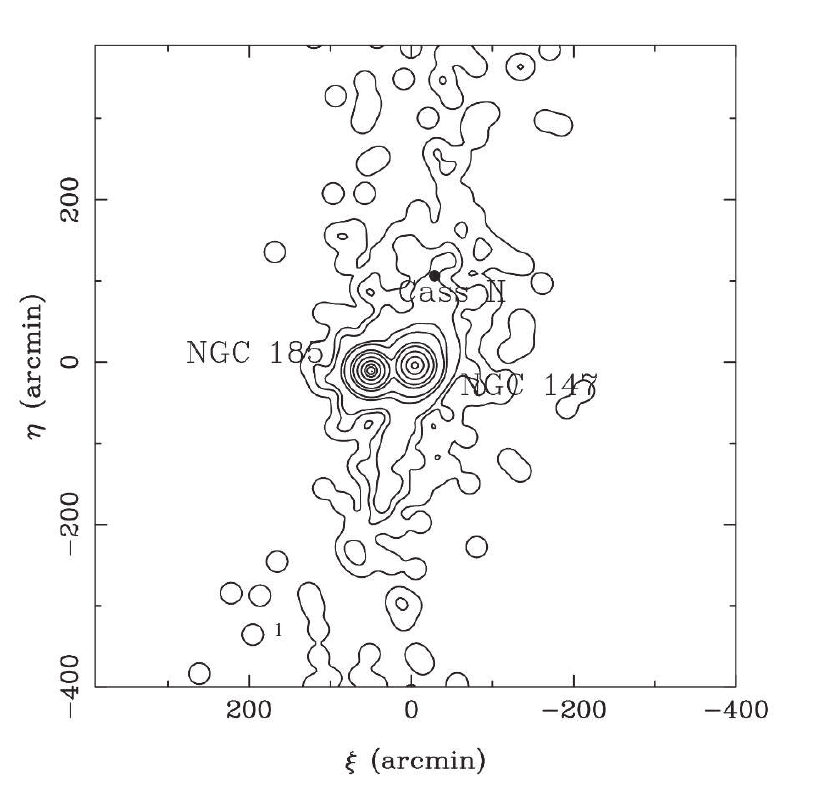}
\caption{Contour plot of the final distribution of stellar particles from the N-Body simulation corresponding to the orbital models shown in Figure \ref{fig:PlaneOrbit}}.
\label{fig:StreamPlaneOrbit}
\end{figure}

As was discovered in 2013, half of the satellites of M31 lie on a very thin plane that seems to be co-rotating \citep{2013ApJ...766..120C,2013Natur.493...62I}. 
NGC~147, NGC~185 and CassII belong to that plane. 
It is therefore interesting to study the orbits of the satellites within this framework. 

From a comparison between the projected orbits of NGC~147, NGC~185 and CassII in the tangential plane and the projected positions of the other satellites that belong to Andromeda plane (see Figure \ref{figOrbits}), it is evident that the orbits that the genetic algorithm finds as the best solutions, do not belong to the Andromeda plane.
In order to constrain the orbits to the Andromeda plane, an extra condition on the satellite's tangential velocities is required. 
In addition to being perpendicular to the line-of-sight velocity, the tangential component needs to be chosen in such a way that the total velocity of each satellite is contained in the Andromeda plane.
We therefore modify the genetic algorithm described in section \ref{secGA}, adding this new constraint on the tangential velocity.
Figure \ref{fig:PlaneOrbit} shows an example of orbital models obtained with this additional constraint. 
{\color{black}Figure \ref{fig:PlaneOrbitXYZ} shows the same orbits projected in the YX plane (left panel), XZ (middle panel) and YZ (right panel), with the positions of the other Andromeda-plane satellites indicated by the black circles. The requirement that the satellites belong to the Andromeda plane implies that the x-component of the velocities stays small. This is clearly seen in the left and middle panels of Figure \ref{fig:PlaneOrbitXYZ}.}
For the model in Figure \ref{fig:PlaneOrbit} and \ref{fig:PlaneOrbitXYZ}, the selected masses are:
$M_{NGC147}=1.02\times{10^{10}}\,{\rm{M_{\odot}}}$,
 $M_{NGC185}=0.64\times{10^{10}}\,{\rm{M_{\odot}}}$,
$M_{CassII}=0.09\times{10^{10}}\,{\rm{M_{\odot}}}$.
The present day tangential velocities (in the M31 frame) are $176 \kms$, $103 \kms$, $88 \kms$  for NGC~147, NGC~185 and CassII respectively. 

We use these parameters to run an additional N-Body simulation (as described in Section \ref{secNbody}) and the final configuration of the stellar particles is presented in Figure \ref{fig:StreamPlaneOrbit}.
{\color{black} Similarly to Model 1 (see figure  \ref{figContourPlots}), the orbits result in a tidal stream in NGC~147 (with a total mass of $7.0\times10^{7}\,\rm{M}_{\odot}$ ), with a smaller, but not negligible, contribution from NGC~185 ($4\times10^{7}\,\rm{M}_{\odot}$). 
However, it is interesting to note that the projected inclination of the stream on the tangential plane in Figure \ref{fig:StreamPlaneOrbit} is different from the observations. 
In general, tidal streams in our simulations tend to align with the orbit of NGC 147 (as shown in figure \ref{figContourPlots}). 
As a consequence, the stream in Figure \ref{fig:StreamPlaneOrbit} appears to be contained in the Andromeda Plane. 
This is not the case for the observed NGC 147 stream shown in Figure \ref{fig_cont} where the direction of the Andromeda plane is depicted by the arrow.}
In a more realistic scenario, the position of tidal streams depends not only on the orbit of the progenitor, but also on its rotation and on the interaction with the host galaxy. 
In follow up work, it will be interesting to expand this analysis and apply the observational information (i.e. metallicity, kinematics) to the NGC~147 stream as constraints on the possible orbits of the system.

\section{Conclusions} 
NGC~147, NGC~185 and CassII measured positions and velocities strongly suggest that these three satellites of the Andromeda galaxy are a subgroup of the Local Group. 
In this work, the results of a statistical analysis show that it is highly unlikely that the similar positions and velocities are due to a chance alignment of the satellites, which further reinforces the sub-group hypothesis. 
Nevertheless, differences in their star formation history and interstellar medium, and the recent discovery of a stellar stream in NGC~147 combined with the lack of tidal features in the other two satellites seemed to contradict the sub-group scenario, since one would expect that the members of a sub-group suffer similar tidal forces and interactions.
However, that is not necessarily the case: using the genetic algorithm we find sets of orbits for which NGC~147 had a closer encounter (and therefore stronger tidal forces) with M31 than the other two satellites of the subgroup. 
This scenario, proposed by Irwin et al. (in preparation) as a possible explanation for the observed differences, was tested in this work using N-body simulations. 
We found that the closer encounter always resulted in the formation of a clear stellar stream in NGC~147, whereas the other two satellites had no significant tidal features. 
We showed that it is therefore possible to have contrasting internal properties for satellites within a subgroup, and that the tidal stream in NGC~147 could be
the result of an interaction with M31 during which the other members of the subgroup were further away and suffered no significant tidal disruption.
\section{Acknowledgements}
M. G. acknowledges the Australia Postgraduate Award (APA) for supporting her PhD candidature and the Astronomical Society of Australia for its travel support. 
N. F. acknowledges the Dean's International Post Graduate Scholarship of the Faculty of Science, University of Sydney. G. F. L. thanks the Australian research council for support through his Future Fellowship (FT100100268), and he and N. F. B. acknowledge support through the Discovery Project (DP110100678).
\bibliographystyle{mn2e}
\bibliography{m31Magda}
\label{lastpage}
\end{document}